\documentstyle[aps,prl,multicol,fancyheadings,epsfig,amsbsy,amssymb,amstex]{revtex}
\def\bbbone{{\mathchoice {\rm 1\mskip-4mu l} {\rm 1\mskip-4mu l}
{\rm 1\mskip-4.5mu l} {\rm 1\mskip-5mu l}}}
\def\bbbc{{\mathchoice {\setbox0=\hbox{$\displaystyle\rm C$}\hbox{\hbox
to0pt{\kern0.4\wd0\vrule height0.9\ht0\hss}\box0}}
{\setbox0=\hbox{$\textstyle\rm C$}\hbox{\hbox
to0pt{\kern0.4\wd0\vrule height0.9\ht0\hss}\box0}}
{\setbox0=\hbox{$\scriptstyle\rm C$}\hbox{\hbox
to0pt{\kern0.4\wd0\vrule height0.9\ht0\hss}\box0}}
{\setbox0=\hbox{$\scriptscriptstyle\rm C$}\hbox{\hbox
to0pt{\kern0.4\wd0\vrule height0.9\ht0\hss}\box0}}}}
\begin{document}
\title{How localized is an extended quantum system ?}
\author{G. Ortiz$^{a}$ and A.A. Aligia$^{b}$}
\address{$^{a}$Theoretical Division, 
Los Alamos National Laboratory, Los Alamos, NM 87545}
\address{$^{b}$Comisi\'{o}n Nacional de Energ{\'{\i }}a At\'{o}mica, 
Centro At\'omico Bariloche and Instituto Balseiro, \\
8400 S.C. de Bariloche, Argentina}

\date{Received \today }

\maketitle

\begin{abstract}
We elaborate on a geometric characterization of the electromagnetic
properties of matter. A fundamental complex quantity, $z_{L}$, is
introduced to study the localization properties of extended quantum
systems. $z_L$, which allows us to discriminate between conducting and
non-conducting thermodynamic phases, has an illuminating physical (and
geometric) interpretation. Its phase can be related to the expectation
value of the position operator (and a Berry phase), while its modulus
is associated with quantum electric polarization fluctuations (and a
quantum metric). We also study the scaling behavior of $z_L$ in the
one-dimensional repulsive Hubbard model. 
\end{abstract}

\vspace*{-0.4cm}
\begin{multicols}{2}

\columnseprule 0pt

\narrowtext
\vspace*{-0.5cm}

{\it Introduction.}
 The continued cross-fertilization between topology and physics has
generated simple ideas that allow one to clarify many subtle issues in
condensed matter and quantum field theories. Several problems in
distinct areas of physics can be phrased in geometrical terms resulting
in a clearer understanding of its structure and a more elegant
expression of its solution. A purpose of this paper is to show that the
concept of localization and the field of quantum phase transitions
belong to that class of problems. Physically, what distinguishes an
insulator from a metal, or superconductor, is the way the system
responds to external electromagnetic probes, e.g., in an insulator a
steady current cannot flow at zero temperature. That macroscopic
characterization of matter translates microscopically into a study of
the localization properties of their quantum states, which differ in a
fundamental way. How localized is an extended quantum system? What are
the topological differences between the states defining a conductor and
an insulator? Can one establish a criterion for localization? are
some of the questions we want to address. The ultimate goal is to
establish the foundations that will allow one to study and predict the 
electromagnetic properties of generic models of interacting quantum
particles from general principles.

The seminal work of Kohn on the nature of the insulating state is a
landmark in the quantum theory of matter \cite{kohn}. Kohn was
apparently the first to point out that the sensitivity of the
ground-state energy, $E_0$, to changes in the boundary conditions is a
measure of the electric static conductivity of  the system, i.e. its
Drude weight $D_c$. In fact, in the thermodynamic limit (TL), he argued
that $D_c$ is finite for a non-disordered conductor (metal or
superconductor), while it vanishes for an insulator. Physically, this
last result is a consequence of the localization properties of the
many-electron wave function, which is exponentially localized in the
insulating case. Almost thirty years later, Scalapino and co-workers
\cite{scala} suggested a complementary criteria to Kohn's. They
proposed to study the limiting behavior of the paramagnetic
current-current correlation function $\Lambda_{xx}$. The appropriate
long-wavelength and low-frequency limits of $\Lambda_{xx}$  should
distinguish between an ideal metal, superconductor and insulator.
Moreover, they speculated that if the system supports a gap one would
expect that $D_c$ equals the superfluid weight. 

Recently, Resta \cite{res,res1} and the present authors \cite{z} have
introduced a complex quantity $z_L$ that allowed one to discriminate
between conducting and non-conducting thermodynamic phases; $|z_L|
\rightarrow 0$ or $1$ in the large system size limit,  depending upon
the system being conducting or insulating, respectively.  $z_L$ has an
enlightening physical interpretation: its phase is related to the
macroscopic electric polarization of matter (Berry phase)
while its modulus provides information on the polarization fluctuations
(related to a quantum metric \cite{marzari,ivo}). 

It is evident that these different indicators are fundamentally
inter-related so, in a sense, they are conceptually complementary to
each other. Then, why another criteria for localization. Apart from 
helping one to understand the nature of
metal(superconducting)-insulator transitions better, some of the
motivations are the following \cite{z}: 
a) physical connection to the modern theory of macroscopic polarization
\cite{Vanderbilt,ort};
b) appealing geometric content; 
c) sharper distinction between conductors and non-conductors (a two value
criteria); 
d) numerically more convenient. 

We have also identified and used geometric concepts to
characterize the internal topology of stable phases of models of
interacting quantum particles. Each topological quantum  number
(related to a Berry phase) labels a thermodynamic state, and
transitions between them  determine the boundaries of the quantum phase
diagram \cite{topo}. We have examined phase stability and the 
electromagnetic properties of various strongly correlated fermion
systems. Here, the  study of superconducting and Mott phase transitions
are among the most  fundamental issues. In a few cases, exact results
have helped to elucidate the nature of these  transitions, but in
general one has to rely on numerical calculations  of finite systems
for which quantities like $D_c$, or  any other correlation function,
vary smoothly at the transition. Consequently, for example, the
boundaries between charge-density-wave (CDW) or spin-density-wave (SDW)
insulators and metallic phases are difficult to establish using
localization indicators, while the boundaries are extremely sharp when
using topological quantum numbers.

In the present manuscript we plan to review these fundamental ideas
from a general perspective, illustrating main concepts in the
context of Hubbard-like models. 

{\it General Framework.}
We will be concerned with generic quantum $N$-body systems
(in a.u. $\hbar=m=e=c=1$) in the Schr\"odinger picture, described by the
non-relativistic Hamiltonian
\begin{equation}
H = \sum_{i=1}^{N} \frac{p_i^2}{2}  + V(\{x_i\}) \ ,
\end{equation}
where, for simplicity (unless otherwise stated), we will assume  a
one-dimensional (1$d$) system with particle position and momentum 
labeled by $x_i$ and $p_i$, respectively.  The generalization to higher
dimensions is straightforward. In condensed matter physics we are
interested in  the behavior of extended systems, i.e., their large-$N$
limit ($N \rightarrow \infty$ and {\it volume} $L \rightarrow \infty$,
such that $N/L = n_0$ is finite). The boundary conditions on the state
$\Psi$ in $L$ then become an essential ingredient. It turns out that
the generalized Bloch boundary conditions provides the system's
configuration space ${\cal M}$ (of volume $L$) with a non-simply
connected structure. Therefore, the state $\Psi$ ($\in {\cal L}_2({\cal
M},N)$) becomes a section of a $U(1)$ fiber-bundle whose base space is
${\cal M}$. In this way, the topologically interesting properties of
the bundle  will be related to the localization properties of the
physical system. 

Consider a finite $N$-particle system and define the complex quantity 
\begin{equation}
z_L = \langle \Psi_0 | \Psi_0(\alpha) \rangle \ , \ | \Psi_0(\alpha)
\rangle = e^{i \alpha \hat{X}} | \Psi_0 \rangle \ ,
\end{equation}
where $\alpha = \frac{2\pi}{L} \ell$ ($n_0=N/L={\rm n}/\ell$, $({\rm
n},\ell)$ are integers with ${\rm n}/\ell$ an irreducible fraction),
$\hat{X} = \sum_{j=1}^N \hat{x}_j$, and $| \Psi_0 \rangle$ is the
ground state ($H  \ | \Psi_0 \rangle  = E_0 \ | \Psi_0 \rangle$,
$\langle \Psi_0 | \Psi_0 \rangle = 1$). Considered as a continuous
function of $\alpha$, $z_L$ plays the role of a characteristic function
generating all moments of the operator $\hat{X}$, and trivially
satisfies $z_L[0]=1$, $|z_L[\alpha]| \leq 1$,
$z_L[\alpha]=z_L^*[-\alpha]$. Note, however, that the operator
$\hat{X}$ is not a genuine operator in the Hilbert space bundle defined
above, although its exponential is a legitimate one. Therefore,
expectation values of arbitrary powers of $\hat{X}$ have only meaning
in terms of $z_L$. Assuming analyticity in the
neighborhood of $\alpha=0$, $z_L$ can be written in terms of cumulants
$C_j(\hat{X})$ 
\begin{equation}
z_L = \exp \left[ \sum_{j=1}^{\infty} \frac{(i \alpha)^j}{j!}
C_j(\hat{X}) \right] \ .
\end{equation}
It turns out that    
\begin{eqnarray}
\alpha^{-1} {\rm Im} \ln z_L[\alpha] &=& \langle \hat{X} \rangle + {\cal O}
(\alpha^2) \ , \nonumber \\
- \alpha^{-2} \ln |z_L[\alpha]|^2 &=& \langle \hat{X}^2 \rangle - \langle
\hat{X} \rangle^2 + {\cal O}(\alpha^2) \ .
\end{eqnarray}
so that the phase of $z_L$ is related to the macroscopic electric
polarization \cite{ort} while its modulus provides information on the
quantum polarization fluctuations \cite{z,topo,ivo}. Kudinov using the
fluctuation-dissipation theorem showed that, in the TL, $\alpha
(\langle \hat{X}^2 \rangle - \langle \hat{X} \rangle^2)$ is finite for
an insulator while it diverges in a conductor \cite{kudi}. 

The analyticity assumption of $z_L$ about $\alpha=0$ in the TL is
equivalent to saying that the system has a gap (see below). In general,
this is true for finite systems, particularly in 1$d$. As we will argue
now, (barring possible pathologies) $z_L \rightarrow 0$ when the system
is a conductor. Basically, the operator $e^{i \alpha \hat{X}}$ shifts
all one-particle wave vectors by $-\alpha$. Therefore, in Fermi or
Luttinger liquids, the Fermi surface is shifted and $z_L$ should vanish
for $L \rightarrow \infty$.  It is also easy to see that $e^{i \alpha
\hat{X}}$ applied to the BCS wave function for the superconducting
ground state converts it into a  current carrying superconducting state
orthogonal to the former in the TL, which also leads to $z_L
\rightarrow 0$. In contrast to the Drude weight, these results are
also  valid in the presence of disorder.

In the case of a system which is insulating due to the effect of
disorder, with correlations playing a secondary role, one expects that
the ground state is a single Slater determinant with the $N$ particles
occupying the $N$ energetically more favorable localized wave
functions. We calculate $z_{L}$ in a more general case, in which the
ground state is a combination of $M$ Slater determinants
\begin{eqnarray}
|\Psi_{0}\rangle &=& \sum_{j=1}^{M} c_j | \Phi_j \rangle \;,\;\;
\mbox{with} \\ \langle \{x_i\} |\Phi_j\rangle &=&
\sum_{P}\frac{(-1)^{P}}{\sqrt{N!}} \psi _{jP1}(x_{1})\psi
_{jP2}(x_{2})...\psi _{jPN}(x_{N}), \nonumber \label{det}
\end{eqnarray}
with $\sum_{j}|c_{j}|^{2}=1$, and the single-particle orbital $l$ of
the $j^{th}$ Slater determinant satisfies
\begin{equation}
|\psi _{jl}(x)| < A \ e^{-|x-R_{jl}|/\lambda }.  \label{bound}
\end{equation}
Then,
\begin{equation}
z_{L}=\sum_{ij}c_{i}^{*}c_{j}\sum_{P}(-1)^{P}\prod_{l=1}^{N}e^{i\alpha
R_{il}} \langle \psi _{il}|e^{i\alpha (x-R_{il})}| \psi_{jPl}\rangle .
\end{equation}
Because of the bound (\ref{bound}), the exponential in the matrix
element can be expanded in powers of $\alpha$  and only the first term
survives in the TL $L\gg 2\pi \ell \lambda $ ($\alpha \lambda \ll 1$).
Thus, orthonormality of the one-particle orbitals leads to 
\begin{equation}
\lim_{L\rightarrow \infty} z_{L} = \sum_{j=1}^{M}|c_{j}|^{2} 
e^{i\alpha \sum_{l} R_{jl}}.  \label{zlim}
\end{equation}
This leads to $|z_{L}|\rightarrow 1$ in the case of only one Slater
determinant, or in the general case provided that $\alpha$ can be
chosen in such a way that the exponents in Eq. (\ref{zlim}) have the
same value for all determinants. For translationally invariant
systems, this is the reason for the introduction of the factor $\ell$
in the definition $\alpha =2\pi \ell/L$ \cite{z}.

For insulating systems in the presence of disorder {\em and}
correlations, we are not able to prove that $|z_{L}|\rightarrow 1$
without introducing additional hypotheses.

On the other hand, under the assumption that $| \Psi_0(\beta) \rangle$
is non-degenerate in the interval $0 < \beta < \alpha=2 \pi \ell/L$, for
any correlated insulator $|z_L| \rightarrow 1$ in the large system size
limit. To understand this statement let's first rewrite $z_L$ in the
following manner   
\begin{equation}
z_L = \frac{2}{N \alpha} \langle \Psi_0 | \hat{P} \Psi_0(\alpha)
\rangle \; , \; \hat{P} = \sum_{i=1}^{N} p_i\ ,
\end{equation}
an expression that is obtained after simple algebraic manipulations, using
the fact that the state $| \Psi_0(\alpha) \rangle$ is an eigenstate of
the $\alpha$-dependent Hamiltonian
\begin{equation}
\bar{H} = H - \alpha \hat{P} \ ,
\end{equation}
with eigenvalue $\bar{E}_0 = E_0 - \alpha^2 N/2$ ($\bar{H} \ |
\Psi_0(\alpha) \rangle = \bar{E}_0 | \Psi_0(\alpha) \rangle$). Assuming 
that this state remains non-degenerate for  $0 < \alpha < 2 \pi \ell/L$, 
one can iterate the equation
\begin{equation}
| \Psi_0(\alpha) \rangle = | \Psi_0 \rangle - \alpha G \left [ \bbbone
- | \Psi_0(\alpha) \rangle \langle \Psi_0 | \right ] \hat{P} |
\Psi_0(\alpha) \rangle \ ,
\end{equation}
with $G = \left ( E_0 - H \right )^{-1}$. This is the starting point of
Rayleigh-Schr\"odinger perturbation theory to any order in $\alpha$. 
For an insulating state the result is 
\begin{equation}
| \Psi_0(\alpha) \rangle = e^{i \gamma} \left \{ | \Psi_0 \rangle -
\alpha \sum_{i \neq 0} \frac{\hat{P}_{i0}}{E_{0i}} | \Psi_i \rangle  +
{\cal O}(\alpha^2) \right \} \ ,
\label{wave}
\end{equation}
where $| \Psi_i \rangle$ are eigenstates of $H$, $\hat{P}_{ij} = \langle
\Psi_i | \hat{P} | \Psi_j \rangle$ and $E_{ij} = E_i -E_j$. 
Notice that since the parameter $\alpha \sim 1/L$, the ${\cal
O}(\alpha^2)$ does not have the usual meaning because the matrix
elements themselves depend upon $L$. However, we argue that
when $L \gg \xi$ ($\xi$ is a correlation length), in the insulating
case, this expansion is well-behaved. Therefore, up to third order in
$\alpha$ (including the normalization of $| \Psi_0(\alpha) \rangle$ to
the same order)
\begin{eqnarray}
&z_L& = e^{i \gamma} \left \{ \left ( 1 -\frac{D_c}{n_0} \right ) \left
( 1 - \frac{3}{2} \alpha^2  \sum_{i \neq 0}
\frac{|\hat{P}_{0i}|^2}{E_{0i}^2} \right ) \right . + \nonumber \\  
&& +
\frac{\alpha^2}{\pi {\rm n}} \sum_{i,j \neq 0}
\frac{\hat{P}_{0i} \hat{P}_{ij} \hat{P}_{j0}}{E_{0i} E_{0j} }   \nonumber \\ 
&& - \left .
\frac{\alpha^3}{\pi {\rm n}} \sum_{l,i,j \neq 0} \frac{\hat{P}_{0l}
\hat{P}_{li} \hat{P}_{ij} \hat{P}_{j0}}{E_{0l} E_{0i} E_{0j} } + {\cal
O}(\alpha^4) \right \} \ ,
\label{zl}
\end{eqnarray}
where the Drude weight (or charge stiffness) is defined as 
\begin{equation}
D_c = \frac{1}{L} \frac{\partial^2 E_0}{\partial \alpha^2} = 
n_0 + \frac{\alpha}{\pi \ell} \sum_{i \neq 0}
\frac{| \hat{P}_{0i} |^2}{E_{0i}} \ . 
\end{equation}
According to Kohn's criteria $\lim_{L \rightarrow \infty} D_c = 0$ in
the insulating state, thus, Eq. (\ref{zl}) is telling us that $\lim_{L
\rightarrow \infty} |z_L| = 1$.  It is useful to bear in mind that Eq.
(\ref{wave}) is only valid if the ground state never becomes degenerate
with any other state \cite{z}. This is an important physical assumption. 

In order to understand the geometric content of the localization
indicator $z_L$, we need to understand the Riemannian structure of our
Hilbert space bundle. Consider a set of normalized states $\{ |
\Psi_0(\mbox{\boldmath{$\alpha$}}) \rangle \}$. Let's assume that this
manifold of quantum states is generated by the action of the group of
Galilean transformations, i.e., 
\begin{equation}
| \Psi_0(\mbox{\boldmath{$\alpha$}}) \rangle = e^{i
\mbox{\boldmath{$\alpha$}} \hat{X}} | \Psi_0(0) \rangle  \;,\;
\mbox{\boldmath{$\alpha$}} \equiv \{ \alpha_{\mu} \}_{\mu=1,d} \ ,
\end{equation}
where $\mbox{\boldmath{$\alpha$}}$ is a vector and  $d$ represents the
number of generators $\hat{X}$ (usually the dimension). What is the
``distance'' between two of these quantum states? To answer this
question, we have to define a (gauge invariant) metric tensor in such a
manifold. 

The problem of establishing a Riemannian structure on an arbitrary 
differentiable manifold of quantum states has been addressed in the 
eighties \cite{metric}. It turns out that a meaningful definition for
the metric is
\begin{equation}
g_{\mu \nu}(\mbox{\boldmath{$\alpha$}}) = {\rm Re} \langle
\partial_{\mu} \Psi_0(\mbox{\boldmath{$\alpha$}}) | \partial_{\nu}
\Psi_0(\mbox{\boldmath{$\alpha$}}) \rangle -
\gamma_{\mu}(\mbox{\boldmath{$\alpha$}})
\gamma_{\nu}(\mbox{\boldmath{$\alpha$}}) \ ,
\end{equation}
where $\gamma_{\mu}(\mbox{\boldmath{$\alpha$}}) = -i \langle
\Psi_0(\mbox{\boldmath{$\alpha$}}) | \partial_{\mu}
\Psi_0(\mbox{\boldmath{$\alpha$}}) \rangle$ is the Berry connection
\cite{ort} and $\partial_{\mu} = \partial/\partial \alpha_{\mu}$. It
can be shown that $g_{\mu \nu}$ is a symmetric and positive definite
second rank tensor. Moreover, what is of interest to us is the fact
that the infinitesimal distance is related to the quantum fluctuations
of the generator $\hat{X}$, i.e., the polarization fluctuations 
\cite{marzari,ivo}
\begin{equation}
g_{\mu \nu}(0) = \langle \hat{X}_{\mu} \hat{X}_{\nu}\rangle - \langle
\hat{X}_{\mu} \rangle \langle \hat{X}_{\nu} \rangle \ ,
\end{equation}
and the expectation values are evaluated over $| \Psi_0(0) \rangle
\equiv | \Psi_0 \rangle$. In a sense, the metric structure on the
manifold is fixed by the quantum fluctuations which determine the 
modulus of $z_L$ in the TL. On the other hand, the antisymmetric
tensor  $\Omega_{\mu \nu} = {\rm Im} \langle \partial_{\mu}
\Psi_0(\mbox{\boldmath{$\alpha$}}) | \partial_{\nu}
\Psi_0(\mbox{\boldmath{$\alpha$}}) \rangle$ plays the role of a
curvature. $\Omega_{\mu \nu}$ is a quantity connected to the
non-dissipative part of the conductance in adiabatic charge transport. 

The Berry phase is an example of a general geometric concept which 
finds realization in several physical problems. It is the anholonomy 
associated with the parallel transport of a vector state in a certain 
parameter space. Anholonomy is a geometric concept related to the
failure of certain variables to return to their original values after
others, driving them, are cyclically changed. In condensed matter, the
charge Berry phase  $\gamma_c$ is a measure of the macroscopic electric
polarization in  band or Mott insulators while the spin Berry phase
$\gamma_s$  
represents the difference between the electric polarization per spin up
and down. $\gamma_c$ is determined by the phase of $z_L$, through the
relation $\gamma_c = \alpha \langle \hat{X} \rangle$.  In systems with
inversion symmetry  $\gamma_c$ and $\gamma_s$ can attain only two
values: 0 or $\pi$  (modulo$(2 \pi)$). Thus, if two thermodynamic
phases differ in the  topological vector $\vec{\gamma} =
(\gamma_c,\gamma_s)$ this sharp difference allows us to unambiguously
identify the transition point even in finite systems. This topological
``order parameter'' was  recently used by us to detect metallic
(superconductor), insulator and conductor-insulator transitions in 1$d$
lattice fermion models \cite{topo}. 

{\it Application to Lattice models.}
To illustrate the concepts developed in previous section, we will study
the localization properties of the well-known 1$d$ Hubbard model at
half-filling ($n_0=1$) \cite{staff}, whose ground state is always
insulating. We concentrate on the analysis of the scaling properties of
$z_{L}$ as a function of the on-site repulsion $U/t$. For the Drude
weight, the scaling $D_{c}\sim \sqrt{L}e^{-L/\xi(U)}$ has been
established by Stafford and Millis \cite{staff}. Using Eq. (\ref{zl}),
the fact that odd terms in $\alpha$ in perturbation theory vanish due
to the inversion symmetry of the model ($\lim_{L \rightarrow \infty}
z_L = \pm 1$), and the additional dependence on $L$ (independent of
$\alpha =2\pi \ell/L$) of the sums entering Eq. (\ref{zl}), one finds
that for large $L$, $|z_{L}|$ scales as $1-\xi /L$, where for large
$U$, $\xi \sim U^{-2}$.

\vspace*{-1.5cm}
\begin{figure}[htb] 
\epsfxsize=25pc 
\centerline{\epsfbox{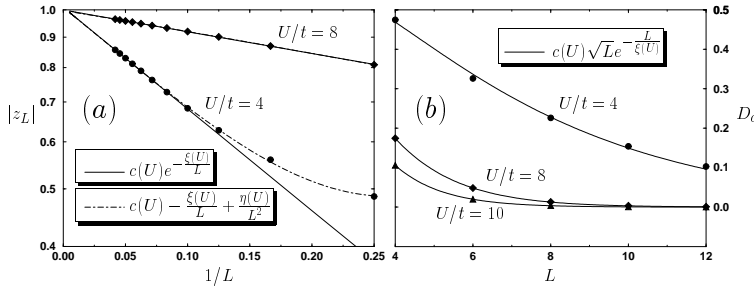}} 
\caption{Scaling behavior of $z_L$ and $D_c$. (a) For the
exponential(parabolic) scaling: $\xi(4)=3.93583(4.00134)$ and
$\xi(8)=0.842294(0.797996)$. (b) $\xi(4)=3.74949(4.0594)$, 
$\xi(8)=1.34926(1.4136)$ and $\xi(10)=1.06704(1.1171)$. We quote in
parentheses their TL values. } 
\label{fig1}
\end{figure}

In Fig. 1a, we display $z_{L}$ for $4\leq L\leq 24$, calculated with
exact diagonalization ($L\leq 12$) and density matrix renormalization
group (DMRG) ($L\geq 12$) for $U/t=4$ and $U/t=8$. For both values, a
parabola in $1/L$ fits very well the results. Moreover, the values of
$z_{L}$ extrapolated to the TL are very close to one (c(4)=1.00979 and
c(8)=0.997598). We also notice that in the scaling regime an
exponential of the form  $|z_{L}|=e^{-\xi /L}$ fits the data very well
(c(4)=1.00993 and c(8)=0.99968), but we do not at present have  any
justification for this functional form. For completeness,  Fig. 1b
shows $D_c$ for $L\leq 12$. 

The ability of $z_{L}$ to detect Mott metal-insulator transitions as a
function of a parameter has been demonstrated in the extended Hubbard
model for infinite on-site repulsion $U/t$ at quarter-filling \cite{z}
and in the Hubbard model with correlated hopping at half-filling
\cite{topo}. In the latter case, due to the pseudospin $SU(2)$ symmetry,
the transition is best determined by the abrupt jump in the charge
Berry phase $\gamma _{c}$. This jump coincides with the change of sign
of $z_{L}$ for $L\rightarrow \infty $ (for any finite system, in
general  $z_{L}\neq 0$, although $|z_{L}|$ can be negligible in the
metallic phase) \cite{z}. In the Hubbard model with correlated hopping,
to extrapolate $z_{L}$ to the TL, it is better to take
$U-U_{c}(L)$ fixed rather than $U$ fixed, where  $U_{c}(L)$ is the
value of $U$ for which  $z_{L}(U_{c})=0$.

{\it Summary.} Over thirty years ago, Kohn \cite{kohn} recognized that
the essential microscopic property that distinguishes the insulating
state of matter is the disconnectedness of its ground-state wave
function. In the present paper, we have shown that one can quantify
this property using concepts borrowed from topology. In this way, we
have seen that the phase of $z_L$ (that is a Berry phase) provides
information on the macroscopic electric polarization of matter, while
its modulus (connected to a quantum metric) measures the degree of
localization in terms of the electric polarization fluctuations. $z_L$
displays a qualitatively different behavior for conductors than for
insulators, thereby providing a two-value criterion to distinguish
between those states of matter. Moreover, we have seen that $z_L$ can
be computed in a simple and very efficient way, using standard
many-body techniques. In particular, we have studied the scaling
behavior of $|z_L|$ for the 1$d$ repulsive Hubbard model in the
insulating phase, and concluded that it is power law in $1/L$. We have
also briefly mentioned the use of topological quantum numbers (the
phase of $z_L$ being one of them) to determine quantum phase diagrams
of interacting particle systems. 

We thank J. Eroles and I. Souza for useful discussions, J. Gubernatis
for a careful reading of the manuscript, and  K. Hallberg and C.D.
Batista for their help with the DMRG calculations. We also thank I.
Souza for pointing out Ref. \cite{metric} to us. G.O. acknowledges
support from DOE. A.A.A. is partially supported by PICT 03-00121-02153
of ANPCyT and PIP 4952/96 of CONICET.

\end{multicols}

\end{document}